\documentclass{article}

\usepackage{natbib}
\usepackage{lineno}

\usepackage{amssymb}
\usepackage{amsmath}
\usepackage{float}
\usepackage{lscape}
\usepackage{longtable}

 \usepackage{graphicx}

\usepackage{amssymb}

\usepackage[skip=10pt]{parskip}

\begin{document}




{\centering

\Large{\textbf{A model of scattered thermal radiation for Venus from 3 to 5 $\mu$m}}\\

\vspace{1cm}

\large{A. Garc\'ia Mu\~noz\footnote{Corresponding author. Email address: tonhingm@gmail.com}}, \\
\small{\textit{Grupo de Ciencias Planetarias, Dpto. de F\'isica Aplicada I, ETS Ingenier\'ia, UPV-EHU, Alameda Urquijo s/n, 48013 Bilbao, Spain;}}\\
\small{\textit{ESA Fellow, ESA/RSSD, ESTEC, 2200 AG Noordwijk, The Netherlands}} 

\vspace{0.5cm}
\large{P. Wolkenberg\footnote{Currently at Space Research Centre of the Polish Academy of Sciences, ul. Bartycka 18A, 00-716 Warsaw, Poland}}, \\
\small{\textit{Grupo de Ciencias Planetarias, Dpto. de F\'isica Aplicada I, ETS Ingenier\'ia, UPV-EHU, Alameda Urquijo s/n, 48013 Bilbao, Spain;}}\\
\small{\textit{ESA Fellow, ESA/RSSD, ESTEC, 2200 AG Noordwijk, The Netherlands}}

\vspace{0.5cm}
\large{A. S\'anchez-Lavega, R. Hueso, and}, \\
\small{\textit{Grupo de Ciencias Planetarias, Dpto. de F\'isica Aplicada I, ETS Ingenier\'ia, UPV-EHU, Alameda Urquijo s/n, 48013 Bilbao, Spain;}}\\
\small{\textit{Unidad Asociada Grupo Ciencias Planetarias UPV/EHU-IAA (CSIC);}}\\

\vspace{0.5cm}
\large{I. Garate-Lopez}.\\
\small{\textit{Grupo de Ciencias Planetarias, Dpto. de F\'isica Aplicada I, ETS Ingenier\'ia, UPV-EHU, Alameda Urquijo s/n, 48013 Bilbao, Spain;}}

\par}				

\vspace{0.5cm}
 \par\noindent\rule{\textwidth}{0.5pt}

\begin{abstract}

Thermal radiation becomes a prominent feature 
in the 
continuum spectrum of Venus 
longwards of $\sim$3 $\mu$m.
The emission is traceable to the upper cloud and haze layers in the
planet's mesosphere.
Venus' thermal radiation spectrum is punctuated by CO$_2$ bands of 
various strengths probing into different atmospheric depths. 
It is thus possible to invert measured spectra of thermal radiation 
to infer atmospheric temperature profiles and offer some
insight into the cloud and haze structure. 
In practice, the retrieval becomes complicated by the fact 
that the outgoing radiation is multiply scattered by the ubiquitous 
aerosol particles before leaving the atmosphere. 
We numerically investigate the radiative transfer problem of 
thermal radiation from the Venus night side between 3 and 5 $\mu$m
with a purpose-built model of Venus' mesosphere.
Special emphasis is laid on the significance of scattering.
The simulations explore the space of model parameters, which includes the 
atmospheric temperature, cloud opacity, and the aerosols' 
size and chemical composition. 
We confirm that aerosol scattering must be taken into account 
in a prospective temperature retrieval, which means an additional complication
to the already ill-posed retrieval problem. We briefly touch upon 
the degeneracy in the spectrum's shape associated with  
parameterization of the Venus clouds.
Reasonable perturbations in the chemical composition and size of aerosols 
do not significantly impact  the model simulations.
Although the experiments are specific to the  
technical characteristics of the 
Visual and Infrared Thermal Imaging Spectrometer
on the Venus Express spacecraft, the conclusions are 
generally valid.
\end{abstract}


\vspace{0.3cm}
\textit{Kew words:} thermal, radiation, Venus, scattering, mesosphere, temperature.




\vspace{0.5cm}
\par\noindent\rule{\textwidth}{0.5pt}
\setcounter{footnote}{0}


\section{\label{intro_sec}Introduction} 

The Venus mesosphere is a complex transition region that extends
from  $\sim$60 to $\sim$100 km above the planet's surface.
It vertically connects the domains of influence of the subsolar-to-antisolar
and retrograde superrotating zonal flow patterns, each dominating the 
global wind dynamics 
above and below the mesosphere, respectively \citep{bougheretal2006}.
Further, the mesosphere is 
key to understand aspects of Venus such as 
the CO$_2$ and sulfur oxidation cycles \citep{millsallen2007}, 
the distribution of the unknown ultraviolet absorber
\citep{titovetal2008}, the deposition of solar energy \citep{crisp1986}, 
and the occurrence of the O$_2$ visible and near-infrared 
airglows \citep{crispetal1996,garciamunozetal2009}.



The thermal radiation emerging 
from the planet's night side provides a valuable window for remotely
investigating the Venus mesosphere. 
Leaving aside the narrow features of thermal emission that occur at specific 
wavelengths 
below 2.3 $\mu$m \citep{allencrawford1984,carlsonetal1991,erardetal2009},
Venus' thermal radiation spectrum commences at $\sim$3 $\mu$m and peaks
somewhere between 10 and 20 $\mu$m. 
The escaping photons originate from within the mesospheric 
upper cloud and haze layers, becoming absorbed and re-scattered
in interactions with the 
aerosol particles. 
Absorption in a few CO$_2$ bands 
makes the thermal radiation spectrum 
amenable to investigation of the mesospheric thermal structure. 
The 
strong CO$_2$ bands at 4.3 and 15 $\mu$m are especially well-suited 
for the purpose, 
and provide a means for sounding the mesosphere from $\sim$100 km 
to $\sim$50--55 km with moderate-resolution spectroscopy 
\citep{carlsonetal1991,grassietal2008,roosseroteetal1995, zasovaetal1999}.
Higher up, the atmosphere becomes too thin   
to leave an imprint on the spectrum,  
whereas lower down aerosol absorption 
prevents the photons from reaching the top of the atmosphere. 

The Venus Express spacecraft of the European Space Agency was set into Venus
orbit in 2006 \citep{svedhemetal2007}.
The Visual and Infrared Thermal Imaging Spectrometer (VIRTIS)
\citep{drossartetal2007,piccionietal2009} 
aboard Venus Express  has since collected a few years worth of spectra. 
The instrument's infrared channel covers the spectral range from  1 to 5 $\mu$m
with a resolving power $\sim$200, which makes of the
instrument a valuable tool for probing the thermal radiation spectrum near 
4.3 $\mu$m. Another instrument on Venus Express, 
the Planetary Fourier Spectrometer \citep{formisanoetal2006}, was meant to probe
both the 4.3 and 15 $\mu$m bands of CO$_2$ at high spectral resolution, but its
unfortunate failure early in the mission impeded the task.



Recent work \citep{grassietal2008,irwinetal2008,leeetal2012} has addressed
various aspects of Venus' thermal radiation spectrum shortwards of 5 $\mu$m 
related to VIRTIS observations.
Although these studies explored the outgoing thermal radiation
under a number of conditions, a systematic sensitivity 
analysis to the relevant parameters in the physical and numerical models
has not yet been presented. 
Thus, the purpose of the current paper is to introduce a radiative transfer 
model (RTM) for simulating the thermal radiation from Venus' night side 
between 3 and 5 $\mu$m, and demonstrate the model capacities
with a number of examples that explore the problem's sensitivity. 
Special attention is paid to the aerosol 
scattering of photons within the atmospheric
medium as aerosols have the capacity of significantly 
modifying the spectrum of outgoing radiation. 
To our knowledge, only one prior work \citep{grassietal2008}
has addressed the details of multiple scattering for temperature retrievals in the
Venus atmosphere. As noted by the authors, computational speed becomes a critical issue
when the multiple-scattering treatment is required in line-by-line calculations.

The temperature retrieval problem in the Venus atmosphere 
differs from the usual treatment for Earth or Mars because multiple scattering is
usually neglected in these two. 
By exploring the relevant space of model parameters, 
the paper intends to facilitate the interpretation of measured spectra 
and their sensitivity.
Subsequent work will address the quantitative characterization and retrieval of
temperatures and aerosol optical properties in the Venus atmosphere.

\section{The forward model}

The RTM solves the radiative transfer equation (RTE) for the atmosphere 
on a line-by-line basis. 
In a preliminary step, the RTM (optionally) produces two libraries of optical 
properties,
one for the gases 
and one for the aerosols.
Later, the RTM invokes the pre-calculated libraries as input into the
RTE solver and outputs the synthetic spectrum. 
In its current version, the model uses DISORT as the RTE solver
for multiple-scattering calculations 
\citep{stamnesetal1988,stamnesetal2000}.\footnote{ftp://climate.gsfc.nasa.gov/pub/wiscombe/Multiple$\_$Scatt/}  
DISORT is a freely-available 
program for monochromatic 
radiative transfer calculations in plane-parallel stratified media that may
include internal and external radiation sources.
Because the interest of the paper lies in modeling Venus' night side,
we focus on the RTM's capacity for treating thermal emission. 
In the calculation of the unscattered component of thermal radiation, 
we integrate the corresponding RTE along the line of sight with a 
routine built for the task.

The methodology for evaluating the gas and aerosol optical properties 
has largely been described elsewhere \citep{garciamunozpalle2011,
garciamunozbramstedt2012,garciamunozmills2012}, 
so that only an overview is presented here.
For the gas, we use the fundamental parameters for position,
shape and strength of transition lines 
contained in the HITRAN 2008 database \citep{rothmanetal2009}. 
Line shapes are generally assumed to be of the Voigt type and approximated 
with simplified expressions \citep{schreier1992}. 
CO$_2$ lines are known to be sub-Lorentzian at distances of tens to hundreds of
wavenumbers \citep{burchetal1969}, a feature that is accounted for by a
$\chi$($\nu$$-$$\nu_0$) function ($\le$1) premultiplying the Voigt line shapes. 
In the RTM, we adopted the parameterization $\chi(\nu-\nu_0)$=1 for 
$|\nu-\nu_0|$$\le$$\nu_{\rm{min}}$, and =$\exp{(-a[|\nu-\nu_0|-\nu_{\rm{min}}]^b)}$
otherwise, with $a$=0.08, $b$=0.8 and $\nu_{\rm{min}}$=5 cm$^{-1}$
\citep{wintersetal1964}. The expression comes from a cell absorption experiment 
carried out at ambient temperature and pressures of up to 5 atm, and is specific
to the blue side of the band. 
Later works investigated the red side of the band, and
whether $\chi(\nu-\nu_0)$ may also depend on
temperature, pressure, or be asymmetric with respect to the band center $\nu_0$ 
\citep{burchetal1969,ledoucenetal1985,menouxetal1987,perrinhartmann1989}. 
It is common, though, that the parameterization $\chi(\nu-\nu_0)$
at shorter wavelengths must be slightly tuned to reproduce actual 
Venus spectra \citep{crisp1986,meadowscrisp1996}.

CO is the only other gas that produces a noticeable signature at the relevant 
wavelengths.
For the broadening of CO lines in CO$_2$, 
we corrected from the air-based parameters in HITRAN according to usual
prescriptions \citep{baileykedziora2012}.
Rayleigh scattering by the CO$_2$ and N$_2$ background gases 
is considered, although its impact is minor with respect to aerosol
scattering. 
The library of gas optical properties samples the pressure direction from 
10 to 10$^{-8}$ bar with four levels for each ten-fold change in pressure, and 
the temperature direction with 27 temperature levels linearly spaced from 140 to 400 K. 
For pressures and temperatures in between, the gas optical properties 
are linearly interpolated.

In this study, five reference temperature profiles were considered, representative of 
average thermal conditions for latitudes of 30, 45, 60, 75 and 85$^{\circ}$, 
and based on data obtained with a number of techniques by the 
Pioneer Venus Orbiter (PVO) and Venera missions
\citep{seiffetal1985}. 
The average PVO profiles 
have been confirmed by recent
radio occultation measurements with the VeRa instrument on Venus Express
\citep{tellmannetal2009}. The profiles reveal that the temperature decays
monotonically with altitude at the low latitudes, but that polewards the 
profiles develop a thermal inversion at 60--70 km altitude. 
The inversion is particularly pronounced in the so-called Venus' cold collar,  
situated at latitudes of $\sim$70$^{\circ}$, but reaches into the polar vortex.
For a given temperature profile, the RTM converts
from altitude to background pressure
by integration of the hydrostatic balance equation.

\begin{figure}[!htbp]
\begin{center}
\vspace{.5cm}
\includegraphics[angle=0,width=1.\textwidth]{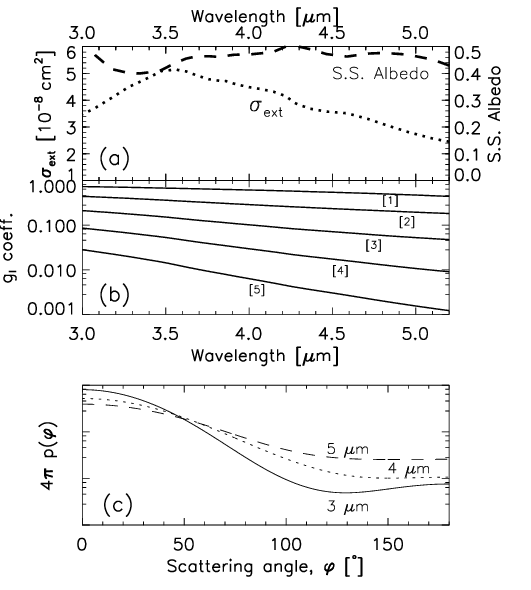}
\caption{ 
\label{haze_fig}}
\end{center}
\end{figure}

The optical properties of the aerosols were calculated from Mie theory 
\citep{mishchenkoetal2002}\footnote{
http://www.giss.nasa.gov/staff/mmishchenko/t$\_$matrix.html}. 
The aerosol sizes were assumed to follow log-normal
distributions described by their effective radius, $r_{\rm{eff}}$,
and effective variance, $v_{\rm{eff}}$. 
In our reference implementation of aerosols, we considered only 
mode-2 particles with $r_{\rm{eff}}$ and $v_{\rm{eff}}$ values of 
1.09 $\mu$m and 0.037, respectively, and a composition of H$_2$SO$_4$:H$_2$O 
in the ratio 84.5:15.5 by mass \citep{molaverdikhanietal2012}.
The adopted real and complex refractive indices are 
laboratory determinations at ambient temperature \citep{palmerwilliams1975}.
The Mie calculations resulted in absorption and
scattering cross sections and in coefficients of the Legendre polynomials
for the scattering phase function over a set of wavelengths. 
At the non-tabulated wavelengths, 
the cross sections and Legendre polynomial coefficients were linearly and
spline interpolated, respectively. For the implemented mode-2 aerosols, 
Fig. \ref{haze_fig} shows 
(a) the extinction cross section and albedo, 
(b) the $g_l$ moments in the DISORT Legendre polynomial expansion for the 
scattering phase function, 
and 
(c) the scattering phase function against the scattering angle 
at a few wavelengths. 

For simplicity, 
the RTM assumes that the aerosol number densities decay with altitude $z$
according to:
$$n_{\rm{aer}}(z)=\exp{(-(z-Z_{\rm{cloud}})/H_{\rm{aer}})} {\big/} {\sigma_{\lambda_{\star}}H_{\rm{aer}}},$$
where $Z_{\rm{cloud}}$ stands for the cloud top altitude, 
$H_{\rm{aer}}$ for the aerosol scale height, and 
$\sigma_{\lambda_{\star}}$= 4.5$\times$10$^{-8}$ cm$^2$ is a reference value 
for the aerosol extinction cross section at $\lambda_{\star}$=4 $\mu$m, see Fig.
\ref{haze_fig}a.
The given law satisfies $\tau_{\rm{nadir}}$=$\int_{Z_{\rm{cloud}}}^{\infty} \sigma_{\lambda_{\star}}
n_{\rm{aer}}(z) dz $= 1, meaning that at 4 $\mu$m a nadir optical thickness of
one is reached at $Z_{\rm{cloud}}$.
More elaborate number density profiles are straightforward to implement.
Given the complexity of the RT problem, though, the above formulation is deemed
appropriate for the present purposes.

\begin{figure}[!htbp]
\begin{center}
\vspace{.5cm}
\includegraphics[angle=0,width=1.\textwidth]{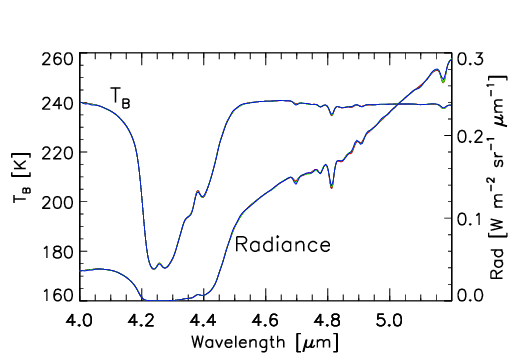}
\caption{ 
\label{resolpower_fig}}
\end{center}
\end{figure}

A major drawback of line-by-line approaches is the computational burden of 
solving the equations over a number of monochromatic bins that may often exceed 
several hundred thousand. 
Our full spectral grid samples the 1800--3500 cm$^{-1}$ interval of wavenumbers
with 1.1$\times$10$^6$ points spaced according to a geometric summation rule.
The summation rule is built from the premise that 
the ratio $\Delta\nu_i/\nu_i$ of bin size and mid-wavenumber within the bin is
constant.  
With bins $\Delta$$\nu$$\sim$(1--2)$\times$10$^{-3}$ cm$^{-1}$, 
this 
ensures 
at least two bins per Doppler width at the core of CO$_2$ lines
throughout the mesosphere. 
However, a relaxation in the size of the spectral grid becomes acceptable 
when moderate spectral resolution is needed in the ultimate model spectrum.
After some trials, we
concluded that undersampling the full spectral grid 
led to negligible errors on the order of 1\% in the model
spectra at the VIRTIS resolution for undersampling factors of up to 50. 
Thus, the calculations presented here utilize a reduced grid that 
samples the 1800--3500 cm$^{-1}$ interval with only
2.2$\times$10$^4$ points. 
After the line-by-line calculation, the spectra are
convolved at the VIRTIS resolution onto a simpler spectral grid. 
For illustration, Fig. \ref{resolpower_fig} shows model spectra for undersampling factors of 
1, 10, 25 and 50. The corresponding curves are nearly indistinguishable.

As a check, we compared our CO$_2$ optical properties in a few conditions of
temperature and pressure with those kindly provided to us by D. Grassi, finding an
excellent agreement between the two sets.
Also, we compared a model spectrum for an aerosol-free atmosphere as calculated by 
D. Grassi's model and our model. Again, the agreement was excellent, even though
the model radiance varied by orders of magnitude.

  
\section{Exploring the outgoing thermal radiation spectrum}

The thermal radiation predicted to emerge from the top of the atmosphere is 
affected by a number of parameters associated with both 
the numerical and physical models. Noting by \textbf{y} the model output, 
\textbf{x} the state vector of model parameters, and 
\textbf{b} an additional set of parameters from the numerical and physical
models,
the forward model \textbf{F} relates them through $\textbf{y= F(x, b)}$ \citep{rodgers2000}. 
The distinction between \textbf{x} and \textbf{b} is subjective, and
simply considers separately the parameters that one might attempt to infer
from a model-observation comparison (\textbf{x}) and those that would
remain fixed in the comparison (\textbf{b}).

In our formulation, 
\textbf{y} is the array of model radiances at the top of
the atmosphere at selected wavelengths. 
Also, it will be convenient to consider that 
\textbf{x}=[$T_0, T_1, ..., T_l, ..., T_L$, $H_{\rm{aer}}$, $Z_{\rm{cloud}}$]
is a vector containing the temperatures $T_l$ at specified altitudes, 
and $H_{\rm{aer}}$ and $Z_{\rm{cloud}}$, the aerosol 
parameters introduced earlier.
Array \textbf{b} contains parameters such as $r_{\rm{eff}}$, $v_{\rm{eff}}$ and the
aerosol chemical composition in the physical model, 
and the number of streams up and down in the RTE solver for the numerical model.  

To explore the impact of each parameter on the RTM's output, 
the concept of weighting function (WF) plays a key role. The WF matrix 
is the matrix of partial derivatives 
$\partial \textbf{F(x, b)} / \partial \textbf{x}$ and expresses the 
output sensitivity to perturbations of the state vector 
about a given position. 
In the framework of the optimal estimation of atmospheric parameters
\citep{rodgers2000}, 
accurate representations of the WF matrix
are critical to speed up the convergence of the inversion algorithm 
and estimate the variance of the inferred parameters \citep{rodgers2000}.
Next, we explore the structure of the WF matrix and assess 
our standard choice for \textbf{b}.

\subsection{Reference spectra for the outgoing thermal radiation} 
 
\begin{figure}[!htbp]
\begin{center}
\includegraphics[angle=0,width=.8\textwidth]{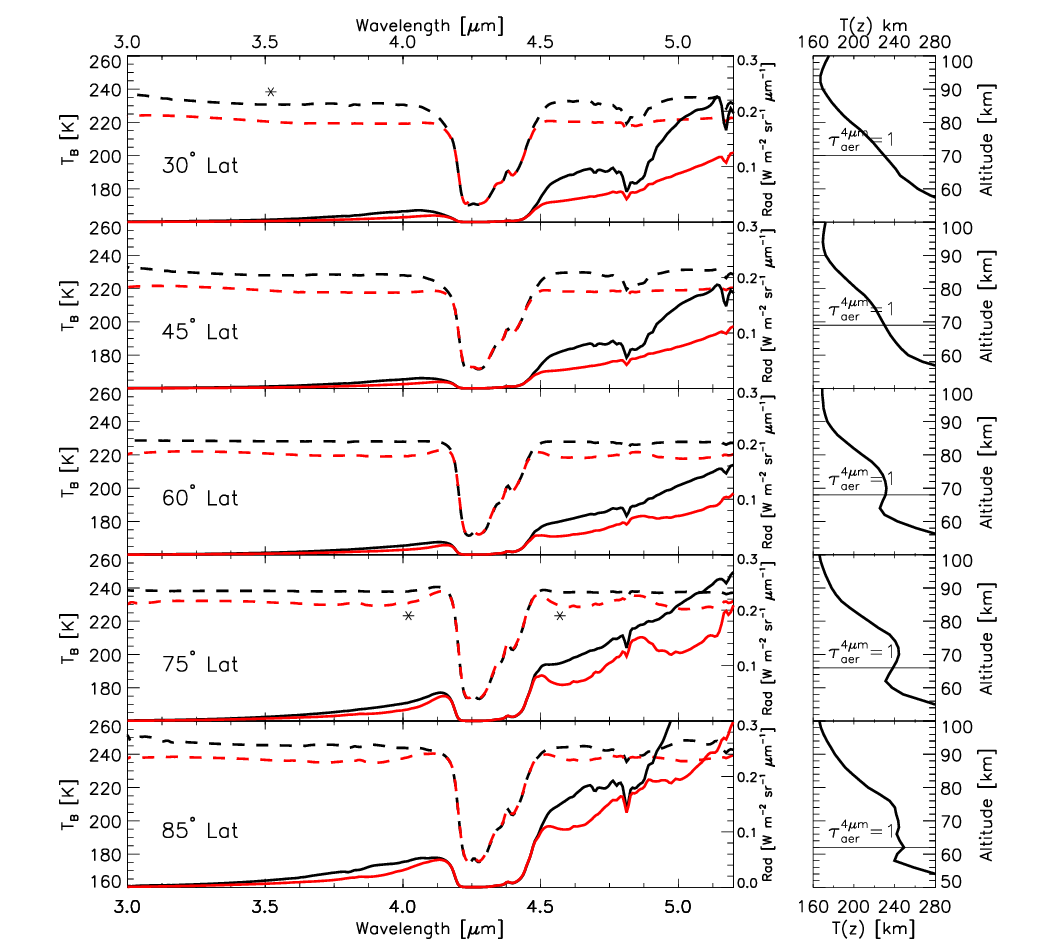}
\caption{ \label{panel_fig}}
\end{center}
\end{figure}

Figure \ref{panel_fig} shows on the left the model spectra for radiance 
(solid) and brightness temperature (dashed) 
for latitudes of 30, 45, 60, 75 and 85$^{\circ}$. 
The brightness temperature $T_{\rm{B}}$ is the temperature of the equivalent 
black body that would produce the measured radiance. 
Panels on the right side show the respective profiles of atmospheric temperature.
We adopted $H_{\rm{aer}}$=4 km in all the simulations, 
but implemented a different $Z_{\rm{cloud}}$ 
for each latitude. 
The prescribed $Z_{\rm{cloud}}$ values are based on Fig. 8a of a recent 
altimetry of the Venus clouds for average conditions 
at 1.6 $\mu$m \citep{ignatievetal2009}. 
In converting the cloud top altitude between the two wavelengths, 
it was estimated from the extinction cross section of Fig. \ref{haze_fig}a, 
that the clouds lie about 4 km lower at 4 $\mu$m than at
1.6 $\mu$m. 
The cloud top level is represented by a horizontal line in the temperature 
graphs. 
To highlight the contribution of multiple scattering, a second set of 
spectra (in red) was produced with the model's scattering option turned off. 



An inspection of the spectra from 4.3 to 5 $\mu$m shows general consistency with 
a similar exploration of the parameter space \citep{leeetal2012}. 
The spectra reveal a number of features.
Generally, $T_{\rm{B}}$ provides a more visual insight into the
atmospheric temperature profiles than the radiance. 
For comparison with observations, though, 
$T_{\rm{B}}$ fails to give a direct measure of the photon counts, especially at
the shorter wavelengths. 

The $T_{\rm{B}}$ spectra for latitudes of 30 and 45$^{\circ}$ show a mild 
bend near 3.5 $\mu$m (marked with the $\ast$ symbol in the top left panel of Fig. \ref{panel_fig}). 
This is the combined effect of the wavelength-dependent cross sections 
for mode-2 particles, see Fig. \ref{haze_fig}a, 
and the monotonic decrease in the atmospheric
temperature profiles up to $\sim$90 km. Since the cross sections 
at 3 $\mu$m are smaller than at 3.5 $\mu$m, 
the spectrum at 3 $\mu$m shows radiance values from deeper 
in the atmosphere than at 3.5 $\mu$m.
The bend is close to impossible to observe in the radiance spectra.
By pointing it out, it becomes clear that the lack of significant structure in
the aerosol cross sections introduces similarly insignificant changes to the
spectra. This imposes a limitation to the capacity 
of models for inferring the aerosol opacity in the 
atmosphere over the 3--5 $\mu$m interval \citep{grassietal2008}. 
In this respect, the strong structure of the aerosol cross sections 
at 15 $\mu$m leads to a comparative advantage for investigating the aerosol
distribution and temperature in the Venus atmosphere \citep{zasovaetal1999}.


Polewards of 45$^{\circ}$, the adopted atmospheric temperature profiles 
exhibit inversions at 60--70 km \citep{tellmannetal2009}. 
Focusing now on the spectrum for 75$^{\circ}$ latitude, 
the shoulders of the 4.3-$\mu$m band  
(marked also with $\ast$ symbols in the second from the bottom left panel of Fig. \ref{panel_fig}) 
mimic to some extent the temperature profile in the inversion region. 
The reason for this is the rapidly-varying opacity of CO$_2$ 
at the edges of the band, that gives gradual access to most of the inversion layer.
The shoulders can also be appreciated, with more difficulties, 
at 60 and 85$^{\circ}$ latitude.

The comparison between the simulations with the full scattering treatment of the
RTM and with the scattering option turned off are particularly revealing. 
Scattering enhances the outgoing radiance and therefore the inferred
brightness temperature. The radiances calculated in multiple-scattering and
non-scattering modes differ by up to a factor of $\sim$2, 
which translates into $T_{\rm{B}}$ differences of
up to $\sim$15 K. 
This is to be expected because the single scattering albedo in the continuum is 
0.4--0.5, and therefore a large fraction of emitted photons undergo various
collisions before being totally absorbed.
Typically, scattering seems to smooth out 
the spectra's structure in the continuum. 
This is apparent in the 60 and 85$^{\circ}$ latitude spectra near 4.3 $\mu$m, 
in which cases the shoulders of the CO$_2$ band 
nearly disappear. 
We tentatively attribute the effect to the weighting functions, broader 
in the multiple scattering case (see below), which mask the inversion layer by combining 
temperatures from a larger range of altitudes.
Interestingly, inside the weaker CO$_2$ bands scattering leads to 
band depths that differ significantly from those obtained 
in the non-scattering mode.
This is clearly seen at 4.7--4.8 $\mu$m because the combined gas-aerosol
albedo remains high enough, and photons can scatter a few times before leaving
the atmosphere.





\subsection{The WF matrix for atmospheric temperature perturbations}

Figure \ref{ms_ns_fig} shows WF matrices for perturbations in the 
temperature of the atmospheric layers, $T_l$. 
The matrices are displayed as $\partial T_B / \partial T_l$ to have them 
dimensionless. 
They were calculated by perturbing sequentially each $T_l$
by 10 K and finite-differencing the calculated brightness temperatures. 
Only the case for 75$^{\circ}$ latitude is presented.
The top and bottom panels show the calculations in the non- and
multiple-scattering modes of the RTM.

Both WF matrices exhibit similar properties. 
In the nadir, 
the probed altitudes range from $\sim$100 km at the strongest absorption 
of the 4.3-$\mu$m
band, to 56--57 km throughout most of the continuum. The latter is a
little more than two scale heights below the prescribed 
$Z_{\rm{cloud}}$ level of 66 km at 4 $\mu$m.
Both of
the CO$_2$ bands at 4.8 and 5.1 $\mu$m leave distinct marks in the WF matrices. 
Generally, at each wavelength the WF can sense the atmosphere over a total 
vertical span of 3--4 scale heights. 
This can be better appreciated in Fig. \ref{maxWF_fig}, that 
shows the location of the maximum 
$\partial T_B / \partial T_l$ at each wavelength for the prescribed
spectral resolution. 

Figure \ref{WFprofiles_fig} displays a few cuts of the WF matrix at selected
wavelengths for the multiple-scattering (black) and non-scattering (red)
calculations.
This figure confirms some of the earlier ideas and shows clearly 
the range of altitudes probed throughout the spectrum. The weighting functions
are notably narrower inside the strong 4.3-$\mu$m band than in the rest of the
spectrum. Typically, sharp weighting functions facilitate the separation of
contributions from different atmospheric altitudes. 
Multiple scattering broadens the weighting functions by 
$\sim$2--3 km. 
Physically, this can be explained by the longer path lengths of 
multiply-scattered photons. 
 

\subsection{The WF matrices for perturbations in the scale height and
cloud top altitude}

Figures \ref{derivH_fig} and \ref{derivZ_fig} show the WF matrices of
derivatives 
$\partial$$T_B$/$\partial$$H_{\rm{aer}}$ and 
$\partial$$T_B$/$\partial$$Z_{\rm{cloud}}$, respectively. 
They were calculated by finite differencing the brightness temperatures at 
a series of $H_{\rm{aer}}$ and $Z_{\rm{cloud}}$ values. 
The calculations are specific to 75$^{\circ}$ latitude and take into account 
multiple scattering. For $H_{\rm{aer}}$=2--6 km and $Z_{\rm{cloud}}$=62--70 km, the
structure of both WF matrices is remarkably similar, 
especially outside the strongest absorption bands. 
Their resemblance may contribute to the degeneracy of the temperature retrieval
problem when both parameters are set independent in the retrieval algorithm 
\citep{grassietal2008}.
Most of the details of the matrices can be understood in terms of the prescribed
temperature profile, and from the fact that 
increasing either $H_{\rm{aer}}$ or $Z_{\rm{cloud}}$ tends to push upwards the 
atmospheric levels contributing to the outgoing radiance. 

\subsection{Sensitivity to other parameters}

Our physical model for mode-2 particles assumes $r_{\rm{eff}}$=1.09 $\mu$m 
and $v_{\rm{eff}}$=0.037, 
and that the aerosols are liquid spherical droplets made of concentrated
H$_2$SO$_4$ in H$_2$O at the 84.5 percent by mass. 
To further our exploration, we investigated different particle sizes
and chemical compositions. 

Typical values quoted for the percentage of H$_2$SO$_4$ in H$_2$O of the aerosols 
are in the range of 75--85\% \citep{crisp1986,grinspoonetal1993,hansenhovenier1974}. 
Figure \ref{chemistry_fig} shows the radiances for our 75$^{\circ}$ latitude atmosphere and
percentage concentrations of 50, 75, 84.5 and 95.6. 
The results show that changes by reasonable amounts
in the aerosol composition affect very moderately the spectra's structure. 

As with the chemical composition, there is some dispersion in the values
reported in the literature for the aerosol size parameters, with 
$r_{\rm{eff}}$ being typically between 1 and 1.2 $\mu$m 
\citep{crisp1986,pollacketal1980}. 
Also, recent works suggest that aerosols in the polar regions might be
made of somewhat larger particles, which might point to differentiated 
cloud formation processes at low and high latitudes 
\citep{barstowetal2012,leeetal2012,wilsonetal2008}. 
Figure \ref{chemistry_fig} shows our calculations with $r_{\rm{eff}}$=1.4 $\mu$m, 
an effective radius appropriate to the larger mode 2' particles found in the
middle clouds \citep{crisp1986}. The calculations show a mild sensitivity
of the spectrum's structure to the aerosol effective radius within accepted
limits for the latter parameter. 

Finally, we tested the impact of the number of streams in DISORT. The
conclusion was that streams of four or more gave nearly identical results. 
Since the computational burden increases rapidly with the number of streams, we
set the parameter to four.

\section{The region from 4.5 to 4.8 $\mu$m}

For conditions of low $Z_{\rm{cloud}}$, the synthetic spectra may 
exhibit a distinct feature near 4.6 $\mu$m. 
This can be appreciated in Fig. \ref{feature_fig}, 
that shows spectra for 75$^{\circ}$ latitude and 
various cloud top altitudes. 
The feature resembles the thermal windows shortwards of 2.3 $\mu$m because, 
like them, 
it matches a local weakening in the CO$_2$ absorption spectrum.
Unlike them, the feature at 4.6 $\mu$m originates in the mesosphere 
and not in the lower atmosphere below 50 km. 
Gas absorption in the affected spectral region is largely dominated by 
the far wings of the strong 4.3-$\mu$m band. 
Thus, the shape and intensity of the feature are affected by 
the parameterization $\chi(\nu-\nu_0)$ for the CO$_2$ wings described earlier.

The impact on the feature at 4.6 $\mu$m
of truncating $\chi(\nu-\nu_0)$ at different distances from the line
center  can be seen in Fig. \ref{cutoff_fig}. 
The calculations were done with the 75$^{\circ}$ latitude temperature profile
but, to emphasize the absorbing properties of the CO$_2$ far wing, we 
set the cloud top at 62 km, somewhat lower than in the standard conditions.
The figure
shows also the result of taking $\chi(\nu-\nu_0)$$\equiv$1 at all wavelengths, 
thus omitting the sub-Lorentzian shape of the CO$_2$ lines. 
It is apparent that long or unattenuated wings tend to annihilate the feature, 
and that adopting $\chi(\nu-\nu_0)$$\equiv$1 alters the blue shoulder of the
4.3-$\mu$m band more than the red one. 
In relation with the latter remark, past works have noted the difficulty of
reproducing the blue shoulder of Venus spectra
\citep{grassietal2008,roosseroteetal1995}, which might require some tuning of the
$\chi(\nu-\nu_0)$ parameterization. 
A conclusive assessment of the
optimal $\chi(\nu-\nu_0)$ parameterization calls for the simultaneous
determination of the temperature field, which is beyond the scope of the current
paper. 

The 4.5--4.8-$\mu$m region is also affected by the 0--1 fundamental band of CO. 
It is beyond the scope of the current paper 
to attempt the retrieval of CO abundances from VIRTIS/VEx thermal emission 
data, a task attempted in a recent work \citep{irwinetal2008}. 
We will, though, comment briefly on the 
impact of multiple scattering on the CO spectral signature. 

Figure \ref{co_fig} shows a few simulations of 
the spectrum from 4.5 to 4.8 $\mu$m with various amounts of CO. 
On this occasion, and to avoid the additional complication of the temperature
inversion, we used the 30$^{\circ}$ latitude temperature profile.
We assumed a constant mixing ratio of CO throughout the mesosphere. The 
upper and lower sets of curves are the simulations for the multiple- and non-scattering
problems, respectively. The conclusion to draw from the comparison is that the
band depth is notably sensitive to whether the simulations are conducted in the
multiple- or non-scattering modes of the RTM.




\section{Some examples from VIRTIS}

Figure \ref{southpole_fig} is a view of Venus' Southern Pole at a few
wavelengths between 3.83 and 5.1 $\mu$m obtained with VIRTIS in orbit \#38
(2006/05/28, cube VI0038$\_$00.CAL). 
The images show the spatial structure of the South Polar vortex
\citep{piccionietal2007}. 
Figure \ref{virtis_fig} shows the corresponding thermal radiation and brightness
temperature spectra at the indicated 
locations. Both representations give complementary insights into the Venus mesospheric
structure. The locations were selected to give a broad perspective of 
thermal conditions and spatial variability in the near-pole region.

The maps at 3.83, 4.60 and 5.1 $\mu$m show a distinct elongated bright pattern,
which is the manifestation of the polar vortex, 
with additional structure within. The level of detail in each monochromatic 
image depends on the actual depth being probed, see Fig. \ref{maxWF_fig}, 
and the signal-to-noise ratio of the data. 
The wavelengths of 4.41 and 4.52 $\mu$m, 
that fall within the main CO$_2$ absorption band,
probe higher altitudes than the other wavelengths. 
For the latter two, the elongated pattern has (nearly) vanished. 
The level of detail of each image is mirrored in the corresponding spectra.

The spectra from the two points in the upper right corner of the images, outside
the vortex, exhibit clear shoulders
at the edges of the CO$_2$ band at 4.3 $\mu$m.
Those shoulders are suggestive of a thermal 
inversion in the cold collar region.
The feature at 4.6 $\mu$m, probably connected with clouds with relatively lower
$Z_{\rm{cloud}}$ values, 
is apparent in a few spectra,
specifically in those that probe the brightest regions in the images. 
Some related structure, see Fig. \ref{feature_fig} for comparison, 
appears also at about 3.9 $\mu$m.




\section{Summary and outlook}

We presented a RTM for scattered thermal radiation in the Venus atmosphere 
from 3 to 5 $\mu$m. 
We explored the RTM's sensitivity to a number of parameters in the physical and
numerical model. 
The exercise demonstrated the potential importance of scattering in the upper cloud and
haze layers over Venus' mesospheric altitudes. In addition, it
served to pinpoint potential difficulties that might occur in the eventual
application of the RTM to inversion problems.
Scattering is rarely represented in RTMs of planetary atmospheres in the thermal
infrared. So, an investigation like this one should help 
clarify some of the
differential characteristics of the propagation of thermal radiation in 
scattering atmospheres. 

We are currently developing an inversion method to infer Venus' mesospheric 
temperatures from VIRTIS spectra. The need to include
multiple scattering means an additional difficulty because multiple-scattering
RTE solvers consume significantly more resources than non-scattering RTE
solvers. This is particularly true if the RTM must evaluate the WF matrix as
part of the inversion algorithm. 
A linearized version of the RTM, 
such that it produces simultaneously 
both the radiance and the WF matrix, would be an interesting avenue to pursue.









\noindent\textbf{Acknowledgements}

AGM and PW acknowledge a postdoctoral fellowship from Gobierno Vasco. 
This work was supported by the Spanish MICIIN project AYA2009-10701 and AYA2012-36666
with FEDER support, Grupos Gobierno Vasco IT-464-07 and UPV/EHU UFI11/55. 
We gratefully acknowledge Davide Grassi for access to some of his model calculations.\\



\bibliographystyle{apalike}
\bibliography{paperclean.bib}

\vspace {1cm}
\textbf{Figure captions}

\textbf{Fig. 1.}

Optical properties of assumed mode-2 particles, $r_{\rm{eff}}$=1.09 $\mu$m 
and $v_{\rm{eff}}$=0.037. 
(a) Extinction cross section and single scattering albedo.
(b) DISORT-type $g_l$ coefficients for the Legendre polynomial expansion 
of the scattering phase function, with $l$ indicating the polynomial degree; 
$g_1$ is the classical asymmetry parameter. 
(c) The phase function against the
scattering angle at selected wavelengths; 
the forward direction corresponds to a zero angle.

\textbf{Fig. 2.}


Model spectra of brightness temperature and radiance degraded at the VIRTIS resolving power. 
Each set of curves includes four different curves, calculated with undersampling factors 
(see text) of 1, 10, 25 and 50. Even at the largest undersampling (50, the value adopted 
for the subsequent model calculations), the accuracy loss is minor.

\textbf{Fig. 3.}

Left. Model spectra of brightness temperature (dashed) and radiance 
(solid) at the top of the atmosphere in down looking. 
Black and red curves 
are for multiple and non-scattering calculations, respectively.
Right. At each reference latitude, 
atmospheric temperature profiles and levels of nadir 
optical opacity at 4 $\mu$m equal to one are shown. 

\textbf{Fig. 4.}

WF matrix, $\partial$$T_B$/$\partial$$T_l$, 
for atmospheric temperature perturbations. Isoline contours with numbers
indicated and drawn in a color code.
The calculations are
specific to the 75$^{\circ}$ latitude temperature profile adopted. 
$Z_{\rm{cloud}}$=66
km at 4 $\mu$m. Top: Non-scattering. Bottom: Multiple scattering.

\textbf{Fig. 5. }

\begin{figure}[!b]
\begin{center}
\includegraphics[angle=0,width=1.\textwidth]{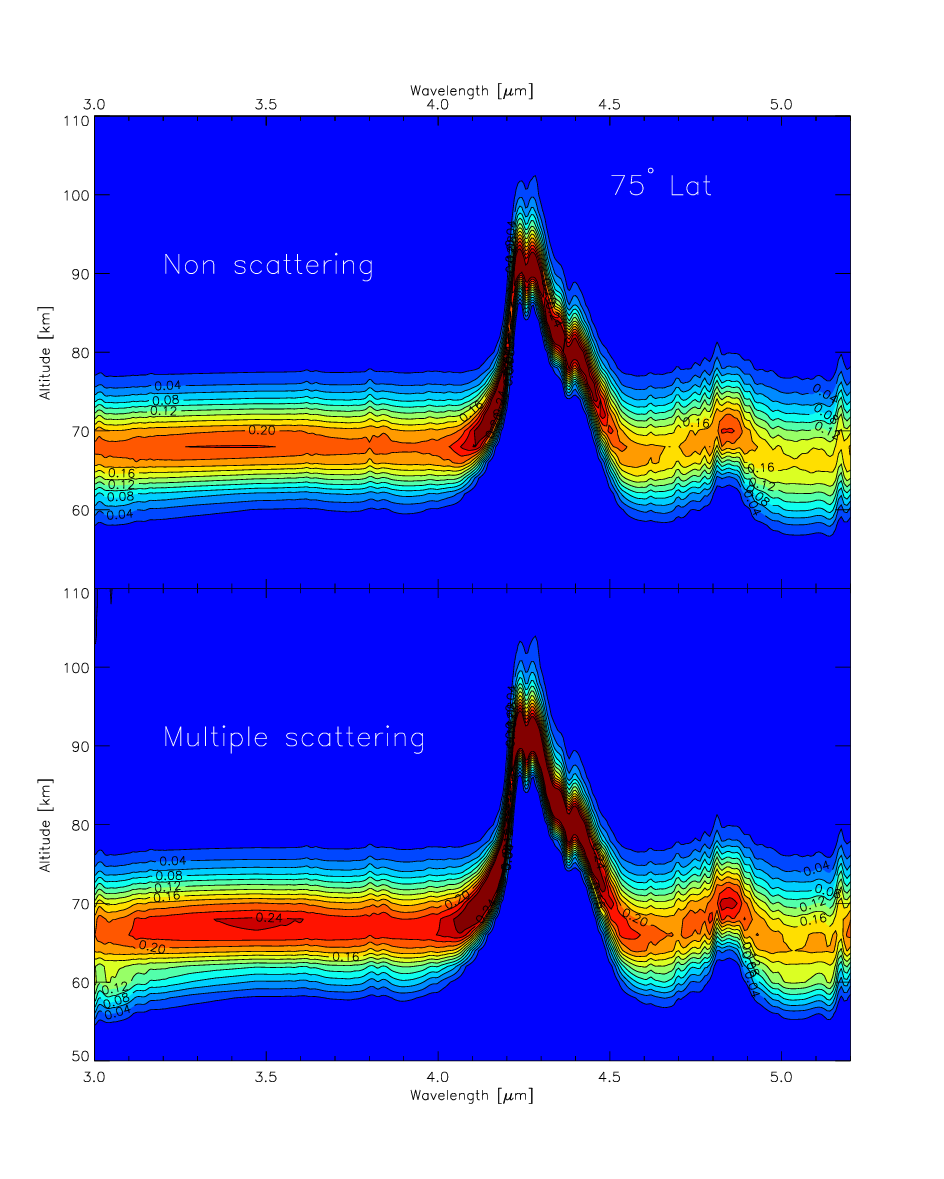}
\caption{ 
\label{ms_ns_fig}}
\end{center}
\end{figure}

\begin{figure}[!t]
\begin{center}
\vspace{1.cm}
\includegraphics[angle=0,width=1.\textwidth]{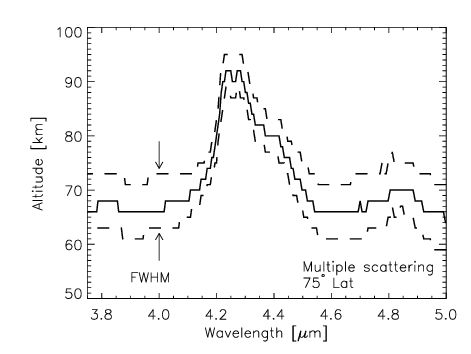}
\caption{ 
\label{maxWF_fig}}
\end{center}
\end{figure}

\begin{figure}[!t]
\begin{center}
\includegraphics[angle=0,width=1.\textwidth]{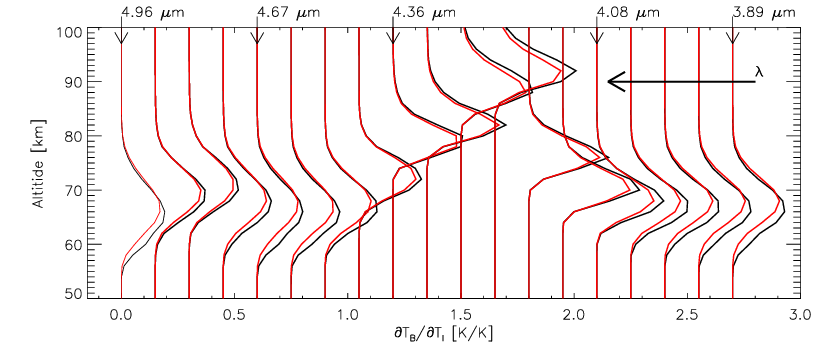}
\caption{
\label{WFprofiles_fig}}
\vspace{-0.5cm}
\end{center}
\end{figure}

\begin{figure}[!htbp]
\begin{center}
\vspace{0.5cm}
\includegraphics[angle=0,width=1.\textwidth]{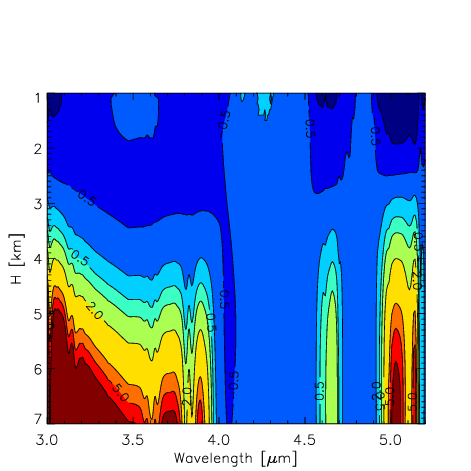}
\caption{ 
\label{derivH_fig}}
\end{center}
\end{figure}

\begin{figure}[!htbp]
\begin{center}
\includegraphics[angle=0,width=1.\textwidth]{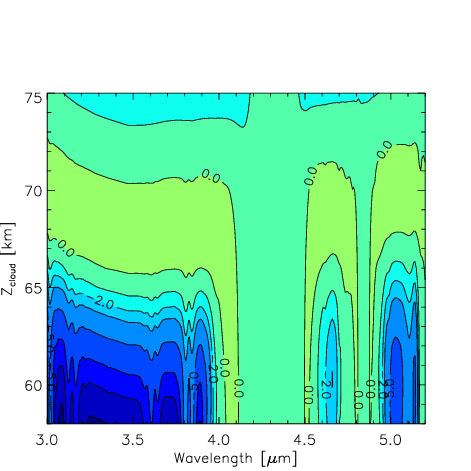}
\caption{ 
\label{derivZ_fig}}
\end{center}
\end{figure}

\begin{figure}[!htbp]
\begin{center}
\vspace{0.5cm}
\includegraphics[angle=0,width=1.\textwidth]{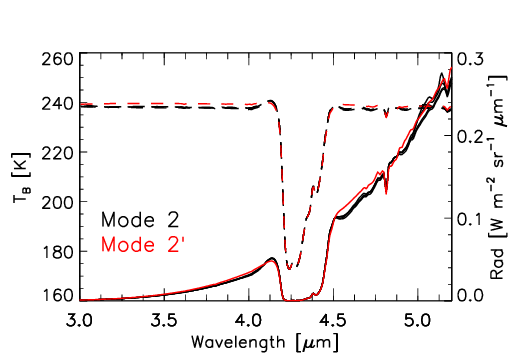}
\caption{   \label{chemistry_fig}}
\end{center}
\end{figure}

\begin{figure}[!htbp]
\begin{center}
\includegraphics[angle=0,width=1.\textwidth]{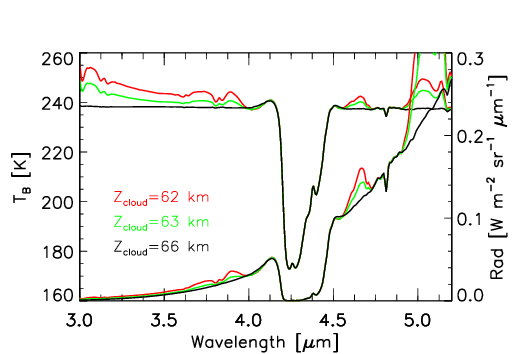}
\caption{ 
\label{feature_fig}}
\end{center}
\end{figure}

\begin{figure}[!htbp]
\begin{center}
\includegraphics[angle=0,width=1.\textwidth]{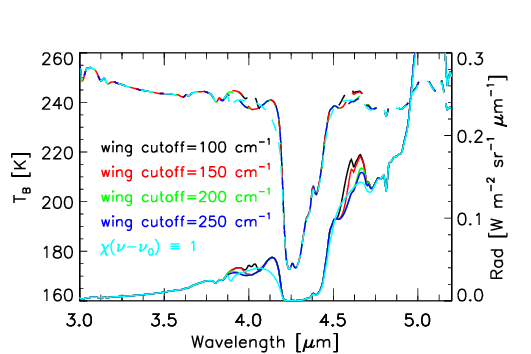}
\caption{ 
\label{cutoff_fig}
}
\end{center}
\end{figure}

\begin{figure}[!htbp]
\begin{center}
\includegraphics[angle=0,width=1.\textwidth]{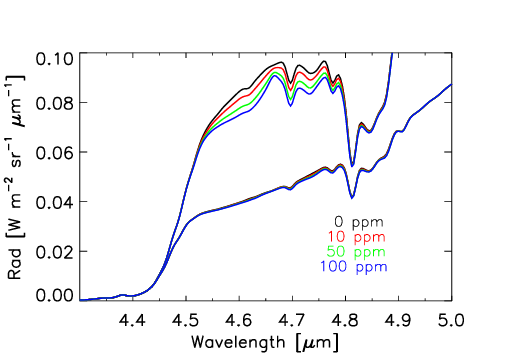}
\caption{
\label{co_fig}}
\end{center}
\end{figure}

Wavelength dependence of the maxima in the WF matrix for atmospheric temperature
perturbations in the multiple-scattering case from Fig. (\ref{ms_ns_fig}).
 
\textbf{Fig. 6.}

From Fig. \ref{ms_ns_fig}, 
weighting functions $\partial$$T_B$/$\partial$$T_l$ at selected wavelengths. 
Black and red curves are for multiple and non-scattering calculations, respectively.
For an easier visualization, the profiles were sequentially shifted by 0.15 K/K. 
From right to left, the wavelengths are 3.89, 3.93, 3.98, 4.03, 4.08, 
4.13, 4.19, 4.24, 4.30, 4.36, 4.42, 4.48, 4.54, 4.61, 4.67, 4.74, 4.81, 
4.88, and 4.96 $\mu$m.

\textbf{Fig. 7.}

WF matrix for perturbations in the scale height of aerosols, $\partial
T_B/\partial H_{\rm{aer}}$, in the multiple-scattering mode of the RTM.
Isoline contours indicated.

\textbf{Fig. 8. }

WF matrix for perturbations in the cloud top altitude, $\partial T_B/\partial
Z_{\rm{cloud}}$,  
in the multiple-scattering mode of the RTM. 
Isoline contours indicated. 

\textbf{Fig. 9. }

Spectra for 75$^{\circ}$ latitude with
alternative descriptions of the size and composition of mesospheric aerosols. 
Dashed line for brightness temperature, and solid line for radiance. 
Black: Curves for $r_{\rm{eff}}$=1.09 $\mu$m and compositions of 50, 75, 84.5 and
95.6 percent H$_2$SO$_4$ in H$_2$O. Red: Curve for $r_{\rm{eff}}$=1.4 $\mu$m 
and composition of 84.5 percent. Calculations in the multiple-scattering mode of
the RTM.

\textbf{Fig. 10. }

Spectra for 75$^{\circ}$ latitude, standard aerosol composition and size, 
multiple scattering, and various
top cloud altitudes. The narrow feature at 4.6 $\mu$m is indicative of low 
$Z_{\rm{cloud}}$ values. 

\textbf{Fig. 11. }

Spectra for 75$^{\circ}$ latitude, $Z_{\rm{cloud}}$=62 km, 
and $\chi (\nu-\nu_0)$ truncated at different distances or, alternatively, 
$\chi (\nu-\nu_0)$$\equiv$1.

\textbf{Fig. 12.}

Spectra for 30$^{\circ}$ latitude and various amounts of well-mixed CO. 
The top and bottom sets of curves are obtained in the 
multiple- and non-scattering modes of the RTM, respectively.

\textbf{Fig. 13.}

VIRTIS view of the Southern Pole in orbit \#38. From top to bottom, wavelengths
of 3.83, 4.41, 4.52, 4.60 and 5.1 $\mu$m.

\textbf{Fig. 14.}

\begin{figure}[!hbp]
\begin{center}
\includegraphics[angle=-90,width=.4\textwidth]{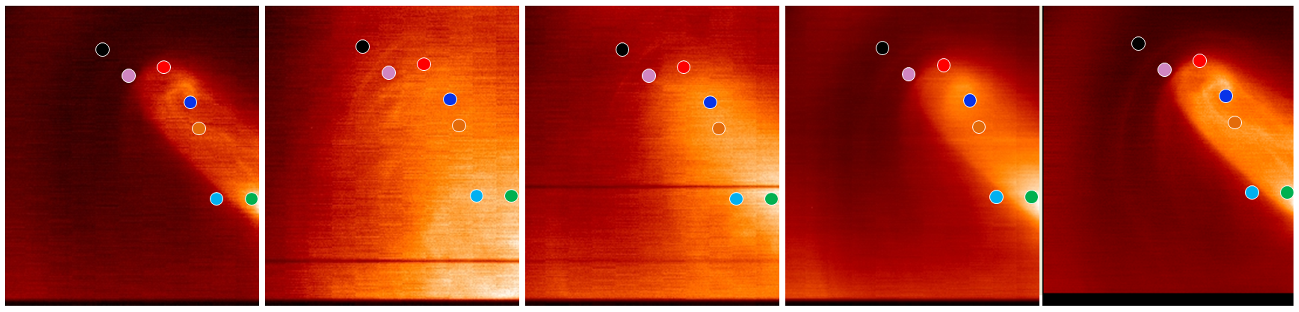}
\vspace{1cm}
\caption{
  \label{southpole_fig}}
\end{center}
\end{figure}

\begin{figure}[!hbp]
\begin{center}
\includegraphics[angle=0,width=1.\textwidth]{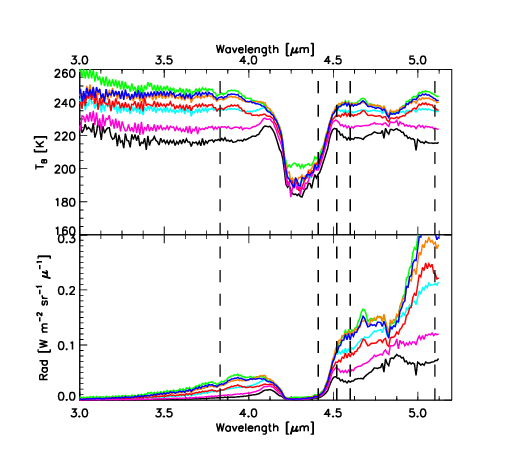}
\vspace{1cm}
\caption{
  \label{virtis_fig}}
\end{center}
\end{figure}

VIRTIS spectra from the color-mark locations in Fig. \ref{southpole_fig}. 
The dashed lines indicate the wavelengths (3.83, 4.41, 4.52, 4.60 and 5.1 $\mu$m
) in the monochromatic images of Fig. \ref{southpole_fig}. 
The nature of the cold collar and the warm vortex is apparent in the structure
of the black and green spectra, respectively.

\textbf{Figures}

\vspace{1cm}

\vspace{1cm}

\end{document}